\documentclass[iop,apj]{emulateapj}

\begin{document}

\submitted{Accepted for publication in The Astrophysical Journal}

\title{Low-frequency oscillations in XTE J1550-564}

\author{Fengyun Rao\altaffilmark{1,2}, Tomaso Belloni\altaffilmark{1}, Luigi Stella\altaffilmark{3}, Shuang Nan Zhang\altaffilmark{4}, Tipei Li\altaffilmark{2,4}}

\altaffiltext{1}{INAF-Osservatorio Astronomico di Brera, Via E. Bianchi 46, I-23807 Merate (LC), Italy}
\altaffiltext{2}{Department of Engineering Physics and Center for Astrophysics,Tsinghua University, Beijing 100084, P. R. China}
\altaffiltext{3}{INAF-Osservatorio Astronomico di Roma, Via di Frascati 33, I-00040 Monteporzio Catone, Rome, Italy}
\altaffiltext{4}{Department of Physics and Center for Astrophysics, Tsinghua University, Beijing 100084, P. R. China}

\shortauthors{Rao et al.}
\shorttitle{Oscillations in XTE J1550-564}

\begin{abstract}
We present the results of timing analysis of the low-frequency Quasi-Periodic Oscillation (QPO) in the Rossi X-Ray Timing Explorer data of the black hole binary XTE J1550--564 during its 1998 outburst. The QPO frequency is observed to vary on timescales between $\sim$100 s and days, correlated with the count rate contribution from the optically thick accretion disk: we studied this correlation and discuss its influence on the QPO width. In all observations, the quality factors ($\nu_0$/FWHM) of the fundamental and second harmonic peaks were observed to be consistent, suggesting that the quasi-periodic nature of the oscillation is due to frequency modulation. In addition to the QPO and its harmonic peaks, a new 1.5$\nu$ component was detected in the power spectra. This component is broad, with a quality factor of $\sim$0.6. From this, we argue what the peak observed at half the QPO frequency, usually referred to as ``sub-harmonic'' could be the fundamental frequency, leading to the sequence 1:2:3:4. We also studied the energy dependence of the timing features and conclude that the two continuum components observed in the power spectrum, although both more intense at high energies, show a different dependence on energy. At low energies, the lowest-frequency component dominates, while at high energies the higher-frequency one has a higher fractional rms. An interplay between these two components was also observed as a function of their characteristic frequency. In this source, the transition between low/hard state and hard-intermediate state appears to be a smooth process.

\end{abstract}


\section{Introduction}

Since the launch of \emph{Rossi X-ray Timing Explorer} (RXTE), an extraordinary progress has been achieved in the knowledge of the variability properties of  black-hole candidates (BHCs) in X-ray binaries. Different types of Quasi-Periodic Oscillations (QPO) have been observed in these systems. While only a few binaries show high-frequency QPOs (HFQPOs, 50--450 Hz, see \citealt{rem06,bel06}), low-frequency QPOs (LFQPOs, mHz to $\sim$10 Hz, see \citealt{rem06,cas05}) are detected in virtually all observed BHCs. As both low and high-frequency QPOs are thought to arise in the accretion flow close to the black hole, the study of their properties and behavior can provide important clues on the physics of accretion onto BHCs.

In the case of LFQPOs, several distinct types showing different properties have been observed. Three main types, dubbed types A, B, and C respectively, stand out in the present scenario. \citet{wij99} and \citet{hom01} first reported type A and type B LFQPO in XTE J1550--564, while \citet{rem02} dubbed the ubiquitous LFQPO that appears together with band-limited noise type-C. The detailed properties of the different types of LFQPOs were investigated by \citet{cas05}. However, the physical difference among types remains unknown.

The X-ray transient XTE J1550--564 was discovered on 1998 September 7th \citep{smi98} with the \emph{RXTE} All Sky Monitor \citep{woo99}. The discovery prompted a follow-up series of almost daily pointed RXTE/PCA (Proportional Counter Array; \citealt{jah96}) observations, which revealed a hard power-law dominated spectrum. Two weeks later, it reached a peak intensity of 6.8 Crab at 2--10 keV. The marked softening of the spectrum during this period indicates a transition from low/hard state (LS) to hard-intermediate state (HIMS), reaching the soft-intermediate state (SIMS) in occasion of the 6.8 Crab peak (see \citealt{hom05,bel09}). After the bright peak, XTE J1550--564 remained in the HIMS for more than three weeks. Strong LFQPOs were observed, with frequency changing quite dramatically during the first $\sim40$ d of the outburst but then stabilizing from day 40--52. The type of QPO also changes at day 40 (from C/C' to B: see \citealt{rem02}). Additional outbursts of the systems followed in 2000, 2002 and 2003.

The optical \citep{ors98} and radio \citep{cam98} counterparts were identified shortly after the discovery of the source. Subsequent optical observations showed that the dynamical mass of the compact object is $10.5\ \pm\ 1.0\ M_\sun$, indicating a black-hole nature. Its binary companion was found to be a low-mass star, and the distance to the source was estimated to be about 5.3 kpc \citep{oro02}. \citet{cor02} discovered a large-scale, relativistically moving and decelerating jet emitting in radio and X-rays.

In this paper, we concentrate on the power density spectra found at the beginning of the 1998 outburst of XTE J1550--564 and analyze in detail the type-C QPO and noise components, focussing on their relative properties. Particular attention is given to the ``harmonic'' peaks, which reveal information about the nature of the observed signal.

\section{Observations and Data Analysis}

\citet{rem02} analyzed the X-ray power density spectra (PDS) for all the 209 \emph{RXTE} observations of XTE J1550--564 during its major outburst of 1998--1999, and detected all three types of LFQPOs. Most of them are of type C and occurred in the first half of the outburst \citep{sob00b}. We reanalyzed the data  and focussed on the 47 observations with type-C QPO between MJD 51,065 and MJD 51,101. We also analyzed all nine observations with type-B LFQPO for comparison.

Custom timing-analysis software under IDL and MATLAB was used. For each observation (In a few cases, an inspection of the PDS showed significant variations in the QPO frequency between different RXTE orbits, which were therefore split.), we produced a set of PDS from 128-s long stretches from the PCA channel band 0--35 (corresponding to 2--13 keV). The time resolution was 1/256 s, corresponding to a Nyquist frequency of 128 Hz. These spectra are then averaged together and logarithmically rebinned. The effect of dead time on Poisson noise was not subtracted directly, but fitted in the power spectra with an additional additive constant. Broad-band and peaked features were modeled by a combination of Lorentzians \citep{now00,bel02} using \emph{XSPEC} v11.3. The power spectra were normalized according to \citet{lea83}; however, the fitting results were converted to squared fractional rms \citep{bel90}.

For each observation, a spectrogram was also accumulated, in order to detect time variations of the QPO parameters within a single observation. This consists in a time sequence of the single 128-s power spectra, i.e. a time-frequency image. We fitted each PDS in the spectrogram (limited to a narrow frequency range centered on the main QPO peak) with a model consisting of a Lorentzian peak and a power law for the local continuum. In this way, we derived the frequency shift with time on a 128-s time scale.

For each observation, we also produced PDS in different channel ranges: eight sub-bands of the 0-35 range plus channels 36--89 (corresponding to 13--33 keV).

\subsection{PDS model}

\begin{figure}
\begin{center}
\includegraphics[height= 8.5cm,width= 8.5cm]{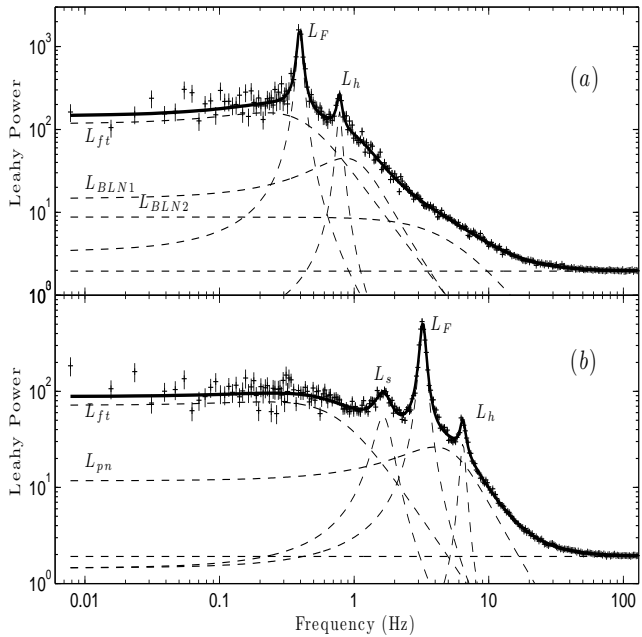}
\caption{Examples of the two types of PDS with type-C QPO. (a) PDS from one of the first 5 observations  (Obs. 30188-06-01-01). The solid line shows the best fit, while the dashed line represents the separate Lorentzian components (see text). (b) PDS of an observation after the first five (Obs.: 30191-01-19-00). In both panels, the solid line shows the best fit, while the dashed line represents the separate Lorentzian components (see text).}\label{fig:f1}
\end{center}
\end{figure}

In order to characterize the LFQPO behavior, a consistent model is required to describe the full PDS for the observations showing a type-C QPO. During the initial rise of the outburst, a state transition from LS into HIMS takes place: the spectrum softens considerably and the QPO frequencies increase well above 1 Hz \citep{cui99,sob00b,rem02}. Therefore, changes are expected in the overall shape of the PDS. We find that the PDS of first 5 observations appear different (see Fig. \ref{fig:f1}). For them, we adopted a model consisting of a flat-top noise component $L_{ft}$, one or two band-limited noise components $L_{BLN1/2}$, a QPO peak $L_F$ with its second harmonic QPO $L_h$ (see Fig. \ref{fig:f1}, panel a). For all other observations with type-C QPOs, we used a model consisting of a flat-top noise component $L_{ft}$, a peaked noise component $L_{pn}$, a QPO $L_F$, with a sub-harmonic $L_s$ and a second harmonic $L_h$ (see Fig. \ref{fig:f1}, panel b). For some PDS a third harmonic appears. These two models fit the data reasonably well, with best-fit reduced $\chi ^{2}$ values less than 2 (for $\sim265$ dof), with a typical value of 1.5. Hereafter, we concentrate on the second part of observations, which we identify as a HIMS. As presented below, the results of our analysis suggest that the fundamental frequency of the oscillation is $L_s$ rather than the conventional $L_F$. However, for clarity, we will refer to $L_F$ as the fundamental throughout the paper. For all noise and QPO components, we consider their characteristic frequency $\nu_{max} = \sqrt{\nu_0^2+(\Delta/2)^2}$, where $\nu_0$ and $\Delta$ are the centroid frequency and the FWHM of the Lorentzian peak respectively (see \citealt{bel02} for a discussion). This is the frequency at which the $\nu P_\nu$ power spectrum peaks. In the case of broad components, this is a more rational choice than the centroid frequency, since the use of Lorentzian models does not have a physical motivation (see \citealt{now00,bel02}). With this definition, it is possible to compare homogeneously narrow and broad features, which have been to evolve from one to the other (see e.g. \citealt{dis01}), and to discover major correlations between characteristic frequencies both in neutron-star and black-hole systems (see \citealt{bel02,van06}). The frequencies of component X, $L_{X}$, will be indicated as $\nu_X$.

For the PDS with a type-B QPO, we used a model consisting of a power-law for red noise, two Lorentzians for the sub-harmonic QPO and the second harmonic QPO, plus a Gaussian for the fundamental QPO (see \citealt{nes03,cas04}). The best-fit reduced $\chi ^{2}$ values were similar to those obtained for  type-C power spectra.

%
%

\section{Results}

\subsection{Correlations between frequencies}

\begin{figure}
\begin{center}
\includegraphics[height= 8.5cm,width= 8.5cm]{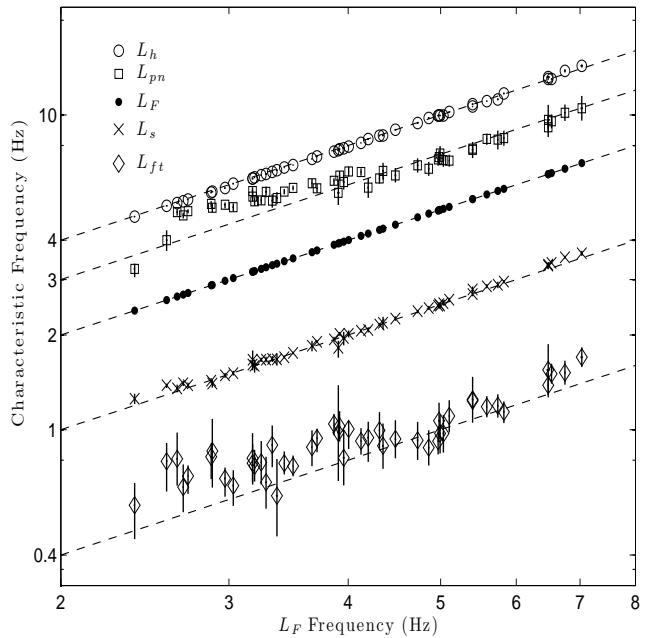}
\caption{Correlations between characteristic frequencies of all Lorentzian components detected in the HIMS as a function of the frequency $\nu_F$ of the $L_F$ component. Circles, squares, filled circles, crosses and diamonds represent the second harmonic $\nu_h$, peaked noise $\nu_{pn}$, fundamental $\nu_F$, sub-harmonic $\nu_s$, flat-top noise $\nu_{ft}$ frequencies respectively (see Fig. \ref{fig:f1}). The dashed lines represent linear correlations with factors 2, 1.5, 1, 1/2 and 1/5 from top to bottom.}\label{fig:f2}
\end{center}
\end{figure}

Following \citet{bel02}, we calculated the characteristic frequency $\nu_{max}$ of each component in the PDS of the HIMS observations. In Fig. \ref{fig:f2}, we plot all the frequencies versus that of the QPO ($L_F$). It is clear that the frequency ratio of the three QPO peaks $\nu_s$, $\nu_F$ and $\nu_h$ is 1/2:1:2, confirming their harmonic relation. The frequency of the flat-top noise component $\nu_{ft}$ appears to to be a factor of $\sim5$ below $\nu_F$, consistent with the correlation reported by \citet{wij99a} and \citet{bel02}. Interestingly, the characteristic frequency of the peaked noise component $\nu_{pn}$ is $\sim1.5$ times $\nu_F$, although some deviation is seen at the low frequency end of the correlation. Ignoring the flat-top noise, whose centroid frequency is zero, assuming  $L_F$ represents the fundamental QPO, the sequence of the peaked components is therefore 0.5--1--1.5--2. Notably, if we take the sub-harmonic peak $L_s$ as the fundamental, the sequence becomes 1--2--3--4. In observations 6 through 11, all with $\nu_F$ around 3 Hz, an additional peak is detected, in three of these observations only at a $3\sigma$ level. Its frequency is consistent with being $3\nu_F$. For these observations, the harmonic sequence would then be 1-2-3-4-6. In Fig. \ref{fig:f2}, some components show systematic deviations from the harmonic lines. At low frequencies they correspond to $L_{pn}$ and $L_{ft}$, i.e. to the broad components in the fit. While $L_{ft}$ is not proposed here to be harmonically related to the others, the deviation of $L_{pn}$ must be discussed. The component is broad and difficult to characterize. In particular, the exactness of the Lorentzian approximation is difficult to evaluate for such component (see Fig. \ref{fig:f1}). This is more the case for the first observations, when the components is at lower frequencies. These deviations could be due to these effects, but their nature and presence must be investigated in other sources.

%
%

\subsection{Frequency evolution}

\begin{figure}
\begin{center}
\includegraphics[height= 8.5cm,width= 8.5cm]{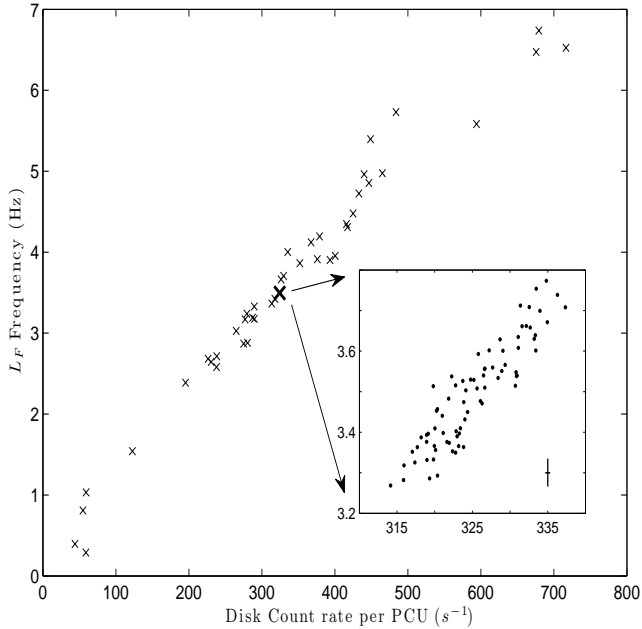}
\caption{Fundamental QPO $L_F$ frequency vs. disk count rate. Each cross corresponds to one observation. The inner panel shows the same relation on 128 s resolution from observation 30191-01-17-00. A typical error bar in frequency is shown in the inner panel.}
\label{fig:f3}
\end{center}
\end{figure}

Although a considerable amount of data showing type-C QPOs have been collected with RXTE, the nature of the frequency variations remains unclear. During the initial rise of the outburst of XTE J1550--564, on average, the type-C QPO between subsequent observations rose monotonically with time and with count rate. \citet{sob00a} showed that the LFQPO frequency is directly correlated with the amount of (unabsorbed) disk flux observed in the 2$\sim$20 keV band. Therefore, we first derived the disk count rate contribution for each observation, using  the total count rate per PCU normalized by disk flux ratio using the best fit models by \citet{sob00b}, assuming the ratio remains constant within one observation. We then produced a frequency vs. disk count rate relation (Fig. \ref{fig:f3}). A linear relation is evident from the figure. For all observations, we produced a similar plot, but with a time resolution of 128 s (the disk rate ratio was assumed to be constant during the observation). The inset in Fig. \ref{fig:f3} shows that the short-time resolution points (limited to one example observation) follow a similar correlation. The linear coefficients of the two correlations (long and short time scale) are inconsistent, most likely because of the choice of a constant disk rate ratio. Notice that correcting the single QPO peaks at 128 s resolution are not compatible with being a coherent oscillation.

%
%

\subsection{QPO width and coherence factor}

\begin{figure}
\begin{center}
\includegraphics[height= 8.5cm,width= 8.5cm]{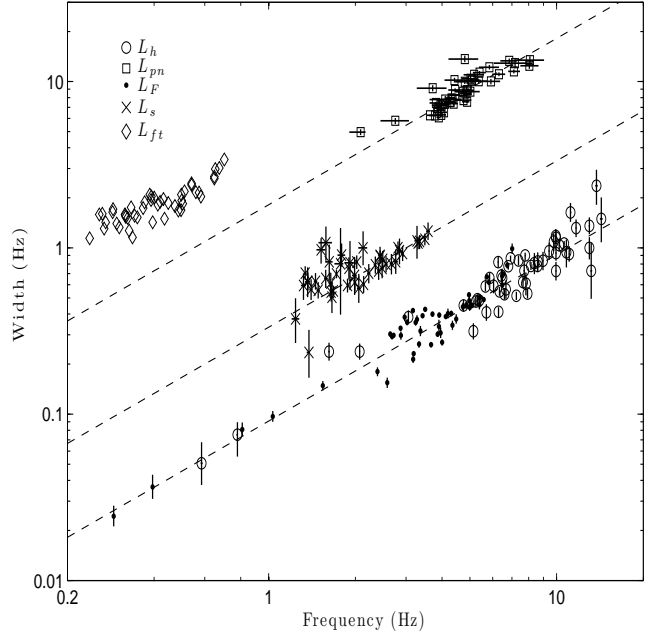}
\caption{Width vs. frequency relation for the peaked Lorentzian components in the PDS, plus the width of the  $L_{ft}$ component plotted as a function of 1/10th of the $L_F$ frequency. Symbols are the same as in Fig. \ref{fig:f2}. The dashed lines indicate Q factors of 0.55, 3 and 11 from top to bottom respectively.}\label{fig:f4}
\end{center}
\end{figure}

It is important to compare the widths of the different QPO peaks in order to uncover the nature of the quasi-periodic signal. In Fig. \ref{fig:f4}, we plot the FWHM of the Lorentzian components vs. their  frequency for all the peaked  components in the PDS. For comparison, for the zero-centered $L_{ft}$ component, we plot their FWHM vs. 1/10 of $\nu_F$. It is clear that for all peaks, the FWHM is positively correlated with the  frequency. The lines in Fig. \ref{fig:f4} indicate constant quality factors $Q=\nu_0/\Delta$. All peaks change their frequency while maintaining approximately the same $Q$. $L_F$ and  $L_h$ have the same $Q$ (around 11), while $L_s$ is consistently broader (Q$\sim$3). The peaked noise component $L_{pn}$ is by definition even broader, but its Q is also consistent with a constant, 0.55, strengthening the identity of this component. Notice that for the peaked components, the characteristic frequency practically coincides with the centroid frequency; the only deviation from this statement is of course the very broad $L_{pn}$.

\begin{figure}
\begin{center}
\includegraphics[height= 8.5cm,width= 8.5cm]{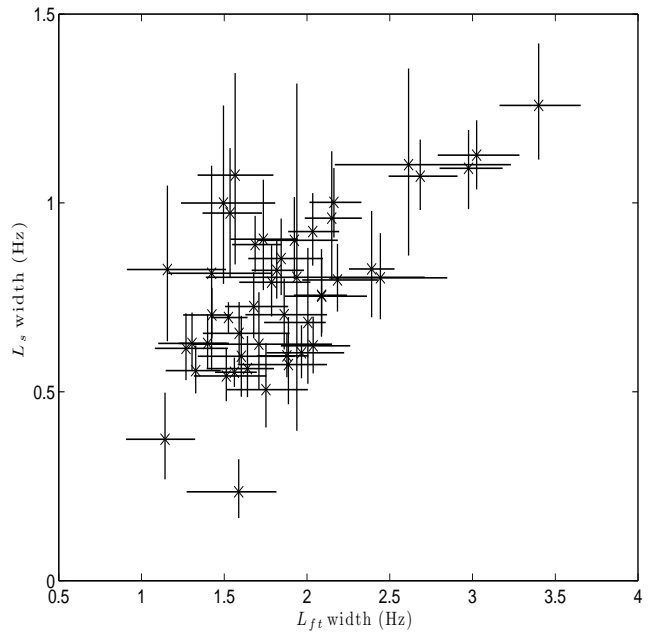}
\caption{Width of the sub-harmonic peak $L_s$ vs. width of the flat-top noise $L_{ft}$ component.}\label{fig:f5}
\end{center}
\end{figure}

The component $L_s$ (subharmonic) is broader than the other peaked components. This cannot be due to frequency modulation, as a different amount of modulation would destroy the harmonic relation on short time scales (the harmonic relation should hold on all time scales). It is however possible that this component undergoes an additional amplitude modulation and the $L_{ft}$ component is the prime candidate for providing it. In order to test it, we plot the correlation between the widths of the $L_s$ and the $L_{ft}$ components in Fig. \ref{fig:f5}. A weak positive correlation is visible. The correlation coefficient is 0.68, corresponding to a chance probability of $1.4\times 10^{-6}$; a linear least-squares fit yields a $R^2$ value of 0.46.

%
%
%
%

\begin{figure}
\begin{center}
\includegraphics[height= 8.5cm,width= 8.5cm]{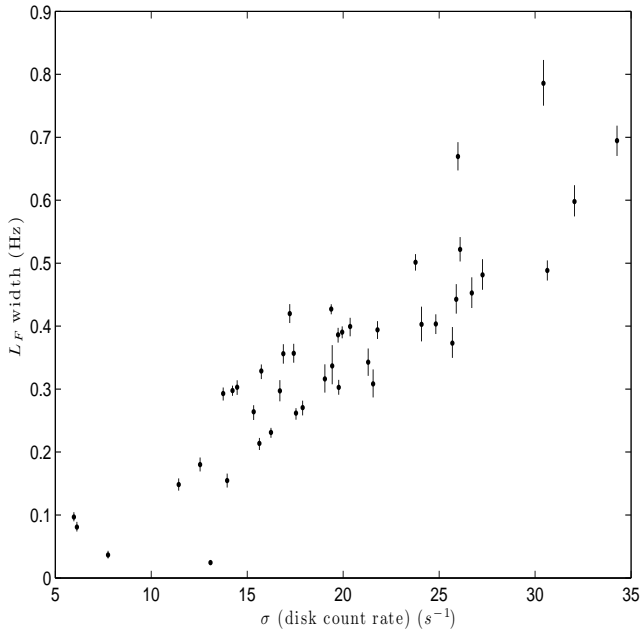}
\caption{Width of the fundamental QPO component $L_F$ vs. standard deviation of disk count rate. Each point represents one observation.
}\label{fig:f6}
\end{center}
\end{figure}

As the QPO fundamental frequency shows a good correlation with disk count rate on $\sim$100 s timescales, it is possible that short-term variations of the centroid frequency due to this correlation are responsible for the width of the QPO peak. In order to estimate the amount of broadening due to this measurable frequency modulation, for each observation we first generated a 1--s binned light curve, then calculated its standard deviation. The resulting linear relation between the width of $L_F$  and this standard deviation is shown in Fig. \ref{fig:f6}. Notice however that the linear coefficient is much larger than that in Fig. \ref{fig:f3}.

%
%

\subsection{Energy dependence}

\begin{figure}
\begin{center}
\includegraphics[height= 8.5cm,width= 8.5cm]{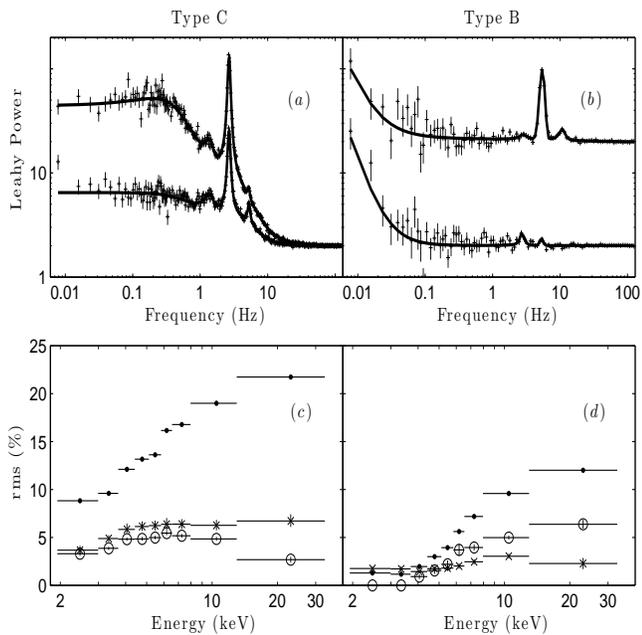}
\caption{Energy dependence for typical examples of type-C (left; Obs.: 30191-01-14-00) and type-B (right; Obs.: 40401-01-58-00). (a,b) PDS for high (13--33 keV; up) and low (2--3 keV; bottom) energies. The solid lines indicate the best fit; In (b), the high-energy PDS is shifted upwards by a factor of 10 for clarity. (c,d) rms dependence of the QPO peaks. The filled circles, open circles and crosses represent the  $L_F$, $L_h$ and $L_s$ components respectively.}\label{fig:f7}
\end{center}
\end{figure}

We performed an energy dependence analysis for all observations with a type-C QPO. Also type-B QPOs were analyzed for comparison. The two types of QPO (and harmonics) show different energy dependences, beyond what shown by \citet{cas05}. An example is shown in Fig. \ref{fig:f7} (this relation is similar for all other observations). In type-C observations, the rms of the fundamental and sub-harmonic peaks increase with energy, while that of the second harmonic increases below a few keV, then decreases at higher-energies. In type-B observations, the rms of the fundamental and harmonic peaks increase with energy, while it is the rms of the sub-harmonic that shows a non-monotonic behavior: it increases below 10 keV and decreases above. A similar relation was shown in Figs. 4 and 5 of \citet{cui99} for type-C QPOs and in Fig. 16 of \citet{hom01} for type-B QPOs. Notice that \citet{cui99} used a slightly different model for the power continuum.

%
%

\subsection{Interplay of components}

\subsubsection{QPOs}

\begin{figure}
\begin{center}
\includegraphics[height= 8.5cm,width= 8.5cm]{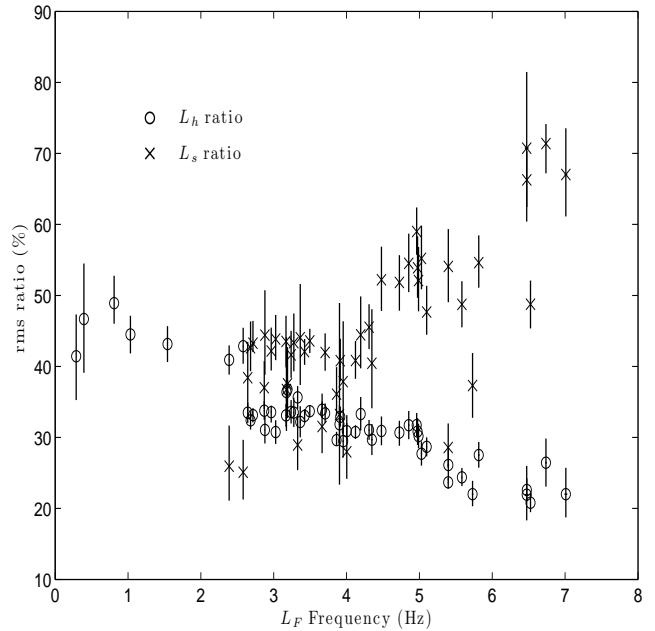}
\caption{Rms ratio (see text) vs. $L_F$ frequency relation for both harmonic and sub-harmonic QPO.}\label{fig:f8}
\end{center}
\end{figure}

In the past, the main focus has gone to the fundamental peak ($L_F$), while the harmonic peaks have received less attention (see e.g. \citealt{sob00a} for XTE J1550-564). We explored the interplay between components, i.e. how they vary differently both with energy and between observations. The energy dependence described in the previous section indicates an interplay, where the harmonic rms ratio decreases with energy. This indicates that the signal becomes more sinusoidal at high energies. We define rms ratio as the rms of the harmonic (or sub-harmonic) peak divided by that of the fundamental. In Fig. \ref{fig:f8} we plot the rms ratio vs. the $L_F$ frequency. It is clear that the harmonic ratio decreases with the $L_F$ frequency, while the sub-harmonic ratio increases, although with large scatter.

%
%

\subsubsection{Broadband components}

\begin{figure}
\begin{center}
\includegraphics[height= 8.5cm,width= 8.5cm]{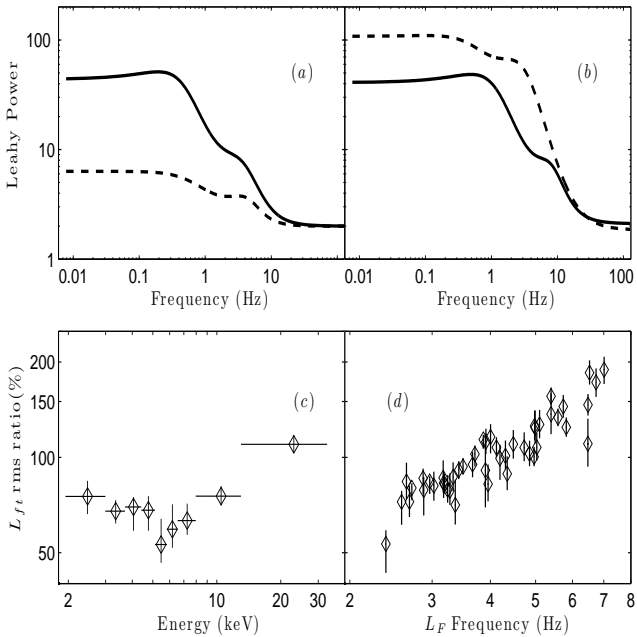}
\caption{(a) best-fit models to power spectra of observation 30191-01-14-00 (solid line: 13--33 keV; dashed line: 2--3 keV). (b) best-fit models to power spectra of observations 30191-01-23-00 (solid line; QPO at 5.6 Hz) and 30188-06-05-00 (dashed line; QPO at 2.93 Hz). (c) energy dependence of the $L_{ft}$ rms ratio (see text) for observation 30191-01-11-00. (d) $L_{ft}$ rms ratio versus $L_F$ frequency for the full set of observations.}\label{fig:f9}
\end{center}
\end{figure}

The same rms-ratio analysis can be performed with broadband components. The $L_{ft}$ rms ratio is defined as the rms of $L_{ft}$ divided by that of $L_{pn}$. We show their dependence on energy and $L_F$ frequency in Fig. \ref{fig:f9}. The $L_{ft}$ ratio increases both with $L_F$ frequency and with energy above 10 keV. Comparing the top panels in  Fig. \ref{fig:f9}, one can see that the overall appearance of the PDS at low and high energies is quite different, as can also be seen in Fig. 4 of Cui et al. (1999). At low energies, $L_{pn}$ is stronger and a break is visible around the QPO peak; at higher energies, $L_{ft}$ dominates and the break can be seen at lower frequencies, consistent with the \citet{wij99a} correlation. This result explains the presence of the second (parallel) correlation, in addition to that of \citet{wij99a}, found by \citet{bel02}, where the characteristic frequency of a broad-component is around the QPO frequency.

%
%

\section{Discussion}

Following \citet{bel02}, we decomposed the PDS in Lorentzian components and found that two different models are needed for the LHS observations (first five) and the HIMS ones. After investigating the interplay among components, we argue that the transition is a smooth process. As the disk flux increases, the $L_F$ frequency increases, the $L_{BLN}$ ratio decreases and then disappear, the $L_{ft}$ and $L_s$ ratio increase, the $L_{pn}$ and $L_h$ ratio decrease. A similar interplay takes place as the energy increase. We can also conclude that $L_{pn}$ and $L_h$ dominate at low $L_F$ frequency, while $L_{ft}$ and $L_s$ dominate at high $L_F$ frequency. All these variations indicate that, depending on energy and time of the observation, the dominant low-frequency component might be $L_{pn}$ or $L_{ft}$, explaining the parallel correlation reported by \citet{bel02}.

Since we don't know the nature of the components, it's difficult to establish the physics of the interplay. However, if we only consider the harmonic ratio, it is clear that the ratio is smaller at high energies and high disk flux, indicating that  the QPO signal becomes more sinusoidal.

We studied the QPO components independently, without assumptions on their harmonic relation. We found that if we include the (broader) $L_{pn}$ component, the frequencies of the four components $L_s$, $L_F$, $L_{pn}$, $L_h$ follow a 1:2:3:4 sequence. This suggests that the $L_s$ component, usually referred to as ``sub-harmonic'', represents the fundamental frequency of the quasi-periodic oscillation. In a few observations, an additional narrow peak at $3\nu_F$ was found. This corresponds to 6 times the subharmonic frequencies, with no peak at 5 times is detected. Notice that there is no priori reason to expect that all harmonics are present in the signal at a detectable intensity. In other words, the absence of a higher harmonic does not weaken the 1-2-3-4 sequence, while the presence of additional peaks at 0.5 and 1.5 times the fundamental are difficult to reconcile with a harmonic series. Of course the detection of a peak at 2.5$\nu_F$ would have strengthened the case. We also showed that the interplay between components can mask the presence of such a $L_s$ peak. One problem with this interpretation is that $L_{pn}$ is much less peaked than the other QPO components, with a $Q$ factor of $\sim$0.55 (see Fig. \ref{fig:f4}). Moreover, at low frequencies, a deviation is observed from 3 times the fundamental. As Fig. \ref{fig:f4} shows, this additional broadening cannot be even larger than what could be caused by $L_{ft}$. A component has been observed to change from being a clear band-limited noise component into a clear QPO or vice versa \citep{dis01}, but it is still difficult to explain the difference between $L_{pn}$ and other QPOs (see below). Alternatively, adopting $L_F$ as fundamental, two peaks would be present at unexpected frequencies: 0.5 (sub-harmonic) and 1.5 times the fundamental. Sub-harmonic signals are observed in nature, but they are associated to non-linear resonance, therefore involving multiple frequencies, adding complication that linear models do not have. As a simple example, an oscillator subject to a periodic force can develop sub-harmonic modes (at $\omega_0/n$) if the period of the force is a multiple of that of the oscillation (see e.g. \citealt{but02,but08}). However, in order to overcome the the effects of damping, energy must be transferred to this mode through non-linear coupling. The presence of a 1.5$\nu_F$ feature would imply another resonance, adding further complications. A simple harmonic series is a linear decomposition of a non-sinusoidal periodic component, without the need of involving non-linear effects. Since its first detections, the possibility that $\nu_s$ represents the fundamental frequency, while sometimes not being detected, has been considered (see \citealt{tak97,bel97,rem02}). However, since no detection of a narrow peak  at $1.5\nu_F$ was found, this assumption, although it could not be excluded, could not be proven.

All frequencies are observed to increase with the count rate associated to the disk component. This is also true on short (128 seconds) time scales, although with different parameters. A similar relation has been observed in type-C QPO of GRS~1915+105 also on short time scales \citep{mar99,mun99}. GRS 1915+105 is a very peculiar object when observed at time scales longer than a second (see \citealt{fen04}) and we can show now that this is also true for another less extreme black-hole transient. The difference between long- and short-time scale correlations could be intrinsic or just caused by error estimation of the disk flux ratio. If the latter is true, the correlation could in principle be used to guide the spectral fitting at short time scales.

The result of the comparison between the FWHM of different QPO peaks is suggestive. The two simplest models for the broadening that gives origin to a QPO peak are amplitude modulation (AM) and frequency modulation (FM). Although these are indistinguishable in the PDS where only one peak is observed, when multiple peaks are present, if the signals are subject to the same modulation, the effect of FM and AM is different. In AM, since the broadening is caused by the windowing due to the varying amplitude, the absolute width of the different peaks will be the same. In FM, the frequency modulation is multiplicative: this translates into peaks that have the same Q factor (see e.g. \citealt{van91,bel97}). Our results are complex. The $L_F$ and $L_h$ components have on average the same Q (notice that the scatter of individual measurements is large, but the overall correlation appears tight and consistent with an equal Q). This suggests that these two signals are frequency modulated. However, the ``sub-harmonic'' component $L_s$ is considerably broader and the $L_{pn}$ at 1.5 times the frequency of the fundamental is even broader. These two components must be subject to additional broadening. It is difficult (although it cannot be excluded) to postulate an additional frequency broadening, since this would break the harmonic relation between frequencies on short time scales. It is more reasonable to assume that these two components are broadened further by an additional amplitude modulation, which affects only these two components, possibly associated to the $L_{ft}$ noise (see Fig. \ref{fig:f2}). Not only we observe that $L_F$ and $L_h$ have the same Q factor. They maintain it throughout the outburst, which implies the modulation effects do not vary with time.

We note that there is a QPO in a black-hole binary which has been analyzed in the time domain, due to strong statistics. It was observed in GRS 1915+105 and it probably of type-C \citep{mor97}. The authors find that the oscillation shows a random walk in phase. Our result is consistent with this, as frequency and phase are directly related: frequency modulation is a type of phase modulation. Unfortunately, our signal is not sufficiently strong to allow such a detailed analysis in the time domain.

\section{Conclusion}

We have analyzed 47 \emph{RXTE} observations with type C LFQPO acquired during the first half of XTE J1550--564 1998 outburst. Satisfactory fits to the power spectra of the 6th--47th observations were obtained with a model consisting of flat-top noise $L_{ft}$, peaked noise $L_{pn}$, fundamental QPO $L_F$, sub-harmonic QPO $L_s$, second harmonic QPO $L_h$ and sometimes a third harmonic QPO components. We identify the peaked noise $L_{pn}$ as a new harmonic component at 1.5$\nu$ the fundamental. This suggests that what we called previously sub-harmonic quasi-periodic oscillation (QPO) may actually be the fundamental.

A similar Q-factor between the fundamental $L_F$ and the harmonic $L_h$ was observed, suggesting a frequency modulation as cause of their width, whereas the sub-harmonic $L_s$ is broader, and the $1.5\nu_{0}$ feature $L_{pn}$ is even broader.

We also found a significant interplay among both QPOs and broad-band components with both frequency and energy. As the disk flux increases, $L_F$ frequency increases, $L_{ft}$ and $L_s$ ratio increase, while $L_{pn}$ and $L_h$ ratio decrease. A similar interplay also takes place as the energy increase. The nature of the interplay remains unknown, since we didn't even know the nature of the components. However, from the interplay we may explain the PDS difference in the first 5 observations, and suggest the transition from LS to HIMS is a smooth process.

\section*{Acknowledgments}

We appreciate very much the insightful comments and helpful suggestions by an anonymous referee. This work was supported by contract PRIN-INAF 2006. SNZ acknowledges partial funding support by the Yangtze Endowment from the Ministry of Education at Tsinghua University, Directional Research Project of the Chinese Academy of Sciences under project No. KJCX2-YW-T03 and by the National Natural Science Foundation of China under grant Nos.10521001, 10733010, 10725313, and by 973 Program of China under grant 2009CB824800.


\end{document}